\RequirePackage{fix-cm}
\documentclass[smallextended]{svjour3}       % onecolumn (second format)
\smartqed  % flush right qed marks, e.g. at end of proof
\usepackage{graphicx}
\usepackage{amssymb}
\usepackage{array}
\usepackage{multirow,bigdelim}
\usepackage{slashbox}
\usepackage{amsmath}
\newcommand{\norm}[1]{\|{#1}\|}

 \def\ket{\rangle}
\journalname{Submitted}
\begin{document}

\title{Duality quantum computer and the efficient quantum simulations \thanks{Project supported by the National Natural
Science Foundation of China (Grant Nos.  11175094 and 91221205), the National Basic Research Program of China
(2011CB9216002).}
}

\titlerunning{Duality Quantum Computing}        % if too long for running head

\author{ Shi-Jie Wei \and Gui-Lu Long}
\institute{Shi-Jie Wei \and Gui-Lu Long \at
              State Key Laboratory of Low-Dimensional Quantum Physics and Department of Physics, Tsinghua University, Beijing 100084, China \\
              \and
              Gui-Lu Long
               \at
              Tsinghua National Laboratory for Information Science and Technology, Beijing 100084, China \\
              \at
              Collaborative Innovation Center of Quantum Matter, Beijing 100084, China\\
              \email{gllong@tsinghua.edu.cn}
}

\date{July 12, 2015}

\maketitle

\begin{abstract}
In this paper, we firstly briefly review the duality quantum computer. Distinctly, the generalized quantum gates, the basic evolution operators in a duality quantum computer are no longer unitary, and they
can be expressed in terms of linear combinations of unitary operators.  All linear bounded operators can be realized in a duality quantum computer, and unitary operators are just the extreme points of the set of generalized quantum gates. A d-slits duality quantum computer can be realized in an ordinary quantum computer with an additional qudit using the duality quantum computing mode.
 Duality quantum computer provides flexibility and clear physical picture in designing quantum algorithms, serving as a useful bridge between quantum and classical algorithms. In this review, we will show that duality quantum computer can simulate quantum systems more efficiently than ordinary quantum computers by providing descriptions of the recent efficient quantum simulation algorithms of Childs et al  [Quantum Information \& Computation,  12(11-12): 901-924 (2012)] for
  the fast simulation of quantum systems with a sparse Hamiltonian, and the quantum simulation algorithm by Berry et al
   [Phys. Rev. Lett. {\bf 114}, 090502 (2015)], which provides  exponential improvement in precision for simulating systems with a sparse Hamiltonian.
\end{abstract}

\section{Introduction}
\label{s1}

One of us, Long, came to know Dr. Brandt first through  his important works in quantum information \cite{brandt1,brandt2,brandt3,brandt4,brandt5,brandt6,brandt7,brandt8,brandt9} and later his role as editor-in-chief of the journal Quantum Information Processing(QIP). Long proposed a new type of quantum computer in 2002  \cite{r1}, which employed the wave-particle duality principle to quantum information processing.  His acquaintance with QIP  began in 2006 through the  work of Stan Gudder who established the mathematical theory of duality quantum computer \cite{r5}, which was accompanied by a different mathematical description of duality quantum computer in the density matrix formalism \cite{r6}.  The development of duality quantum computer owes a great deal to QIP， first in the term of Dr. David Cory  as editor-in-chief, and then the term of  Dr. H. E. Brandt as the editor-in-chief. For example, the zero-wave function paradox was pointed out firstly by Gudder \cite{r5}, and  two possible solutions were given in Refs. \cite{qiudw} and   \cite{cuijx}.   Long has actively participated in the work of QIP as a reviewer when Dr. Brandt was the editor-in-chief, and as a member of the editorial board from 2014. At this special occasion, it is our great honor to present a survey of the duality quantum computer in this special issue dedicated to the memory of Dr. Howard E. Brandt.

As is well-known, a moving quantum object passing through a double-slit behaves like both waves and particles. The duality  computer, or duality quantum computer exploits the wave-particle duality of quantum systems  \cite{r1}. It has been proven that a moving $n$-qubit duality computer passing through a $d$-slits can be perfectly simulated by an ordinary quantum computer with $n$-qubit and an additional $d$ levels qudit \cite{longijtp,r3,r4}. So we do not need to build a moving quantum computer device which is very difficult to realize. This also indicates that we can perform duality quantum computing in an ordinary quantum computer, in the so-called duality quantum computing mode \cite{r3,r4}.
There have been intensive interests in the  theory of duality
computer in recent years
  \cite{r1,r5,r6,qiudw,cuijx,longijtp,r3,r4,rcaohx,r8,r7,r9,rcaohx2,rlongrev1,rlongrev2,lichunyan,longliuyang,factor,r10pp,N4,liuy15},
and  experimental studies have also been reported \cite{haol11,zhengc13}.

This article is organized as follows.  In section \ref{s2}, we briefly describe the generalized quantum gate, the divider and combiner operations.    Section \ref{s3} reviews the  duality quantum computing mode, which enables the implementation of duality quantum computing  in an ordinary quantum computer.
In section \ref{s4}, we outline the main results of mathematical theory of duality quantum computer.  In section \ref{s5},  we give the duality quantum computer description of  the work of Childs et al  \cite{w1} which simulates a quantum system with sparse Hamiltonian efficiently. In section \ref{s6}, we give the duality quantum computer description of  the  work of Berry et al  \cite{Berry} which simulates a quantum system having a sparse Hamiltonian with exponential improvement in the precision. In section \ref{s7}, we give a brief summary.

\section{Duality quantum computer, Divider, Combiner operations}
\label{s2}
\begin{figure}
\begin{center}
\includegraphics[width=9cm,angle=0]{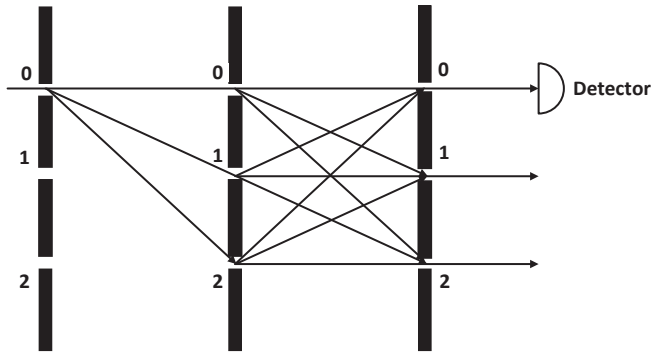}
\caption{An illustrated picture for a three-slits duality quantum computer. The input is from the 0-th slit, and the output of duality quantum computing is taken from only 0th-slit  on the right wall  \cite{longijtp}.\label{f1}}
\end{center}
\end{figure}

A duality quantum computer is a moving quantum computer passing through a
$d$-slits\cite{r1}.  In Fig. \ref{f1}, we give an illustration for a duality
quantum computer with 3-slits \cite{longijtp}.  The quantum wave starts from the 0-th
slit in leftmost wall, and then goes to the middle screen with three
slits (this is the divider operation).  Between the middle screen and
the rightmost screen, some unitary operations are performed on the
sub-waves at different slits.  They are then collected at the
0-slit in the rightmost screen, and this is the quantum wave combiner operation. The result is then read-out by the detector placed at the 0-slit on the right wall.

In a duality quantum computer, the two new  operations are the quantum wave
divider (QWD) operation and quantum wave combiner (QWC) operation
  \cite{r1}.  The QWD operation divides the wave
function into many identical parts with reduced amplitudes.  The
physical picture of this division is simple and natural : a quantum system passing
through a $d$-slits with its wave function being divided into $d$
sub-waves. Each of the sub-waves has the same internal wave function and are different
only in the center of mass motion characterized by the slit.  Conversely, the
combiner operation adds up all the sub-waves into one wave function.
It should be noted that one divides the wave function of the same quantum system into many parts in quantum divider, whereas in quantum cloning one copies the state of one
quantum system onto another quantum system \cite{noclone1,noclone2}(which may also holds true for classical systems\cite{noclone3}). So, the division operation
does not violate the non-cloning theorem.

Considering a quantum wave divider corresponding to a quantum system passing
through a $d$-slits. Writing the direct sum of Hilbert space $\oplus^{d-1}_{i=0}H_i$ as the form of $H^{\oplus^ d}$ where $H_i=H$, $i\in \lbrace 0,\cdots,{d-1}\rbrace $. The divider structure characteristics  which describes the properties of a quantum wave
divider can be denoted as $\{ p_i, i= 0,\cdots,{d-1}\}$ where each
 $p_i$ is a complex number with a module less than 1 and satisfy  $\sum_{i=1}^d |p_i|^2=1$. The divider operator $D_m$ which  maps $H\rightarrow H^{\oplus^ d}$ is defined by
   \begin{eqnarray}
 D_m\psi=\oplus_{i=1}^{d-1} (p_i\psi_{i}). \label{edm}
 \end{eqnarray}
 where $\psi_i=\psi$, $i\in \lbrace 0,\cdots,{d-1}\rbrace $.
This is the most general form of the divider operator, and  it describes a general multi-slits. In a special case, the multi-slits are $d$ identical slits, then $p_i=\sqrt{1/d}$.

The corresponding combiner operation $C_m$  can be defined as follows
\begin{eqnarray}
C_m(\psi_0\oplus \cdots\oplus \psi_{d-1})=\sum_{i=0}^{d-1} q_i
\psi_i,\label{ecm}
\end{eqnarray}
where $ q=\{q_0,\cdots,q_{d-1}\}$ denotes the combiner
structure that describes the properties of a quantum wave combiner.
Each $q_i$ is a complex number that satisfy the module less than 1, and
$\sum_i |q_i|^2=1$.  In the case of $q_i=\sqrt{1/d}$,  the combiner structure is  uniform.

Now, we consider the uniform divider and combiner structures which correspond to $p_i=\sqrt{1/d}$ and $q_i=\sqrt{1/d}$, respectively. In this case, the combined
operations of divider and combiner leave the state unchanged. The process can be described as follows
\begin{eqnarray}
C_mD_m\psi=C_m\oplus_{i=1}^d \sqrt{1/
d}\;\psi_i= \sum_i{1/d }\;\psi=\psi.
\end{eqnarray}
If the divider structure and combiner structure satisfy certain relation, this property also holds. The details will be given in the next section.

It will be shown later in the next section that the divider and combiner
structure $D_m$ and $C_m$ can be expressed by a column or a row of elements
of a unitary matrix respectively.  For duality quantum gates with the form of
$L_r$ in Eq.(\ref{e3}), the relation of the two unitary matrices makes the
structures of  $C_m$ and $D_m$  adjoint of each other.

\section{Duality Quantum Computing Mode in a Quantum Computer }
\label{s3}
The most general form of duality quantum gates has been given
in Refs.  \cite{r3,r4}.  For the convenience of readers, we use the expressions from duality quantum computing mode \cite{r3,r4,rlongrev1}.
Compared to ordinary quantum computer where only unitary operators are allowed, the duality quantum computer offers an additional capability in information processing:  one can perform different gate operations on the sub-wave functions at different slits   \cite{r1}. This is called the duality parallelism, and it enables the duality quantum computer to perform non-unitary gate operations.
The generalized quantum gate, or duality gate is defined as follows
\begin{eqnarray}
L_c=\sum_{i=0}^{d-1} c_i U_i,\label{e1}
\end{eqnarray}
where $U_i$ is unitary and $c_i$ is a complex number and satisfies
\begin{eqnarray}
\sum_{i=0}^{d-1}| c_i|\le 1. \label{e2}
\end{eqnarray}
The duality quantum gate is called real duality gate or real generalized quantum gate when it is restricted to positive real $c_i$. In this case, $c_i$ is denoted by $r_i$, and they are constrained by the condition of $\sum_i r_i=1$. The real duality gate is denoted as $L_r$. So, the real duality quantum gate can be rewritten as
\begin{eqnarray}
L_r=\sum_{i=0}^{d-1} r_i U_i. \label{e3}
\end{eqnarray}
This corresponds to a physical picture of an
asymmetric $d$-slits, and $r_i$ is the probability that the duality
computer system passes through the $i$-th slit.

According to the definition of duality quantum gates, they are generally non-unitary. It naturally provides the capability  to perform non-unitary evolutions. For instance, dynamic evolutions in open quantum systems should be simulated in such machines.  More interestingly, it is an important issue to study the computing capabilities of duality quantum computing. An important step toward this direction is that  Wang, Du and Dou   \cite{r8} proposed an theorem which limits what can not be a duality gate in a Hilbert space with infinite degrees of freedom.

\begin{figure}
\begin{center}
\includegraphics[width=9cm,angle=0]{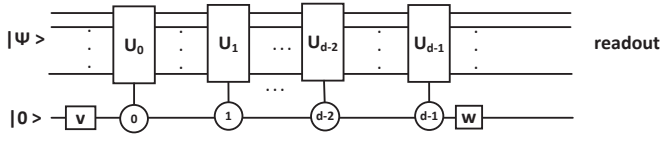}
\caption{  The quantum circuit of duality computing in a quantum computer.  $|\Psi\rangle$ denotes the initial state of duality quantum computer with $|0\rangle$ as the controlling  auxiliary  qudit.  The squares represent unitary operations and the circles represent the state of the controlling qudit,   Unitary
operations $U_{0}$, $U_{1}\cdots$, $U_{d-1}$ are activated only when
the qudit holds the respective values indicated in
circles \cite{longijtp}. \label{f2}}
\end{center}
\end{figure}

The divider operation can be expressed by a general unitary operation $V$ and
the combiner operation can be expressed by another general  unitary operation $W$.
The two unitary operations are implemented on an auxiliary qudit which represents a $d$-slits.  The quantum
circuit of duality quantum computer is shown in Fig. \ref{f2}.
There are $d$ controlled operations between the operations ${V}$ and ${W}$. The $d$ energy levels of the qudit represent  the $d$-slits.

We divide the duality computing processing into four steps to reveal the computing theory in a quantum computer.

Step one: Preparing the initial quantum system  $|\Psi\ket|0\ket$ where  $|\Psi\ket $ is the initial state. Then performing the divider operation by implement the $V$ on the auxiliary qudit, and this operation transforms the initial state into
\begin{eqnarray}
|\Psi\ket|0\ket\rightarrow |\Psi\ket
V|0\ket=\sum_{i=0}^{d-1}V_{i0}|\Psi\ket|i\ket.
\end{eqnarray}
 $V_{i0}$ is the first column element of the unitary matrix $V$ representing the coefficient in each slit. $ V_{i0}=p_i, $ denotes the divider structure. Note that  $V_{i0}$ is a complex number with $|V_{i0}|\le 1$ and satisfies
$\sum_{i=0}^{d-1}|V_{i0}|^2=1$. So $V$ is a generalized quantum division operation.

 Step two: We perform the qudit controlled operations $U_{0}$, $U_{1}\cdots$, $U_{d-1}$ on the target state $|\Psi\ket $ which leads to  the following transformation
\begin{eqnarray}
\sum_{i=0}^{d-1}V_{i0} U_i|\Psi\ket |i\ket.
\end{eqnarray}
This corresponds to the physical picture that implements  unitary operations simultaneously on the sub-waves at different slits.

Step three: We combine the wave functions by  performing the unitary operation $W$. The following state is obtained,
\begin{eqnarray}
\sum_i V_{i0} U_i|\Psi\ket W|i\ket.
\end{eqnarray}

Step four: Detecting  the final wave function when the qudit is in state
$|0\ket$ by  placing a detector at slit 0 as  shown in Fig. \ref{f2}. The wave function becomes
\begin{eqnarray}
\sum_{i} W_{0i}V_{i0}U_i|\Psi\ket|0\ket=\sum_{i}
(W_{0i}V_{i0})U_i|\Psi\ket|0\ket.
\end{eqnarray}
It should be noted that $W$ is a generalized quantum  combiner operation and $ q_i=W_{0i}, $ which is the combiner structure in Eq. (\ref{ecm}).
Hence $c_i=W_{0i}V_{i0}$ is the coefficient in the generalized  duality gate defined in Eq. (\ref{e1}). Now, we have successfully realized the duality quantum computing in an ordinary quantum computer.

Considering  a special case that $W=V^{\dagger}$, the coefficients
$r_i=V^{\dagger}_{0i}V_{i0}=|V_{i0}|^2$  satisfy
\begin{eqnarray}
\sum_i r_i=\sum_i|V_{i0}|^2=1,
\end{eqnarray}
where $r_i$ is defined in Eq. (\ref{e3}) corresponding to the real duality gate $L_r$.

Generally speaking, $c_i=W_{0i}V_{i0}$ is a complex number.  The sum of $c_i$'s can be denoted as
\begin{eqnarray}
\sum_i c_i=\sum_i W_{0i}V_{i0}=(WV)_{00}. \label{eqvw}
\end{eqnarray}
The value of $(WV)_{00}$ is just  an element of a unitary
matrix $ WV $, and naturally has the constraint $|(WV)_{00}|\le 1$.  Hence the
most general form of duality gates allowable by the principles of quantum
mechanics is the form of (\ref{e1})

%where $c_i$ are complex numbers with module less or equal to 1, and
%with the constraint $|\sum_i c_i|\le 1$ and can be written as the
%product of elements of two unitary matrices, namely
%$c_i=W_{0i}V_{i0}$ .
In a recent study, Zhang et al  \cite{rcaohx2} has proved that it is
realizable and necessary to decompose a generalized quantum gate  in terms of two unitary
operators $V$ and $W$ in Eq.  (\ref{eqvw}) if and only if the coefficients
satisfy $ \sum_i |c_i|\le 1. $
Obtaining the explicit form of the decomposition is a crucial step  in duality quantum algorithm design and related studies.

\section{Mathematical theory of duality quantum computer}\label{s4}

The mathematical theory of duality quantum computer has been the subject of
many recent studies  \cite{r1,r5,r6,r3,r4,rcaohx,r8,r7,r9,rcaohx2}.  Here we briefly review the mathematical description in duality quantum computing.  In this case, the mathematical theory of the
divider and combiner operations are restricted to a real structure,
namely each $p_i (i= 0, 1, \cdots, {d-1} )$ is real and
positive, and the uniform structure is also a special case of the
 combiner structure.  The following results are from Ref. \cite{r5} and
we label the corresponding lemma, theorem and corollaries by a
letter G, and the corresponding operators are labeled with a
subscript $p$.

Here are the properties of generalized quantum gates and related operators \cite{longijtp}.

Defining the set of generalized quantum gates on $ H $  as $\mathcal{G}(H)$  which  turns out to be a convex set. Then we have

{\bf  Theorem G 4. 1} The identity $I_H$ is an extreme point of
$\mathcal{G}(H)$, where  $I_H$ is the identity operator on $H$.

Any unitary operator is in $\mathcal {G(H)}$ and $I_\mathcal{H}\in
\mathcal{G(H)}$.   The identity $I_H$ is an extreme point of
$\mathcal{G}(H)$ which indicates  $\sum_i p_i
U_i=I_{\mathcal{H}}$ if and only if $U_i=I_{\mathcal{H}}$ for all
$i$.

{\bf  Corollary G 4. 2} The extreme points of $\mathcal{G(H)}$ are
precisely the unitary operators in $\mathcal{H}$.

We can conclude from {\bf Theorem G 4. 1 } and {\bf Corollary G 4. 2 } that the ordinary quantum computer is included in the duality quantum computer.

Denoting $\mathcal{B(H)}$ by the set of bounded linear operators on
$\mathcal{H}$ and let $\mathbb{R}^+\mathcal {G(H)}$ be the positive
cone generated by $ \mathcal {G(H)}$.  That is
\begin{eqnarray}
\mathbb{R}^+\mathcal {G(H)}=\{\alpha A: A\in \mathcal
{G(H)},\alpha\ge 0\}.
\end{eqnarray}

{\bf Theorem G 4. 3} If dim$ \mathcal{H}< \infty$, then
$\mathcal{B(H)}=\mathbb{R}^+\mathcal {G(H)}$.

This theorem shows us that the duality quantum computer is able to simulate
any operator in a Hilbert space $\mathcal{H}$ if dim$ \mathcal{H}< \infty$.

It should be pointed out that these lemmas, corollary and theorems also hold for divider and combiner with a general complex structure \cite{r1}.   One limitation has been given explicitly by Wang, Du and Dou \cite{r8} that what can not be a generalized quantum gate when the dimension is infinite. It is an interesting direction to study the computing ability of duality quantum computer in terms of this theorem.

\section{  Description of Childs-Wiebe Algorithm for Simulating Hamiltonians in a Duality Quantum Computer}\label{s5}

Simulating physics with quantum computers is the original motivation of Richard Feynman to propose the idea of
quantum computer  \cite{Fey82}. Benioff has constructed a microscopic quantum mechanical model of computers as
represented by Turing machines \cite{Be}.  Quantum simulation is apparently unrealistic using classical computers,
 but quantum computers are naturally suited to this task.  Simulating the time evolution of quantum systems or the dynamics of quantum systems is a major potential application of quantum computers.
Quantum computers accelerate the integer factorization problem exponentially through the use of Shor algorithm \cite{shor}, and the unsorted database search problem in a square-root manner through the Grover's algorithm \cite{grover} (see also the improved quantum search algorithms with certainty \cite{longalg,toyama}). Quantum computers can simulate quantum systems exponentially fast \cite{Fey82}.
  Lloyd proposed
the original approach to quantum simulation of time-independent local
Hamiltonians based on product formulas  \cite{Lloyd} which attracted many attentions \cite{lulong,sor,ch,ah,flong.flong2}. However, in this formalism,  high-order approximations
lead to sharply increased algorithmic complexity, the performance of simulation algorithms based on product formulas is limited \cite{w1}. For instance,  the Lie-Trotter-Suzuki formulas, which is high-order product formulas, yields a new efficient
approach to approximate the time evolution using a product of unitary operations whose length scales exponentially with the order of the formula  \cite{su}.  In contrast, classical methods based on  multi-product formulas require a sum of unitary operations  only in polynomially scales to achieve the same accuracy  \cite{classic}.  However, due to the unclosed property of unitary operations  under addition, these classical methods cannot be directly implemented on a quantum
computer.

The duality quantum computer can be used as a bridge to transform classical algorithms
in to quantum computing algorithms. Duality parallelism in the  duality quantum computer enables us to perform the non-unitary operations.  Moreover,duality quantum  gate has the form $L_c=\sum_{i=0}^{d-1} c_i U_i\label{e1}$. This is the linear combinations of unitary operations. Duality quantum computer is naturally suitable for the simulation algorithms of Hamiltonians based on multi-product formulas.

Childs and Wiebe proposed a new approach to simulate Hamiltonian dynamics based on implementing linear combinations of unitary operations\cite{w1,w2}.  The resulting algorithm has superior performance to existing simulation algorithms based on product formulas and is optimal among a large class of methods. Their main results are as follows

 \begin{theorem}\label{thm:mainresult}
Let the system Hamiltonian be $H=\sum_{j=1}^{m} H_{j}$ where each $H_{j}\in C^{2^n\times 2^n}$
is Hermitian and satisfies $\parallel H_{j}\parallel\leq h$ for a given constant $h$.  Then the Hamiltonian evolution $e^{-iHt}$ can be simulated on a quantum computer with failure probability and error at most $\epsilon$ as a product of linear combinations of unitary operators.  In the limit of large $m,ht,1/\epsilon$, this simulation uses
\begin{equation}\label{eq:mainresult}
\tilde O\left(m^2 h te^{1. 6\sqrt{\log(m h t/\epsilon)}}\right)
\end{equation}
elementary operations and exponentials of the $H_{j}$s.
\end{theorem}

Considering this simulation algorithm is based on implementing linear combinations of unitary operations, it can be implemented  by duality quantum computer. Now, we give the duality quantum computer description of this simulation algorithm.

The evolution operator $U(t)  $ satisfies the Schr{\"o}dinger equation

\label{eq:schrodinger}
\begin{equation}
i\frac{d}{dt}U(t)=HU(t),
\end{equation}
and  time evolution operator can be formally expressed as $ U(t)=e^{-iHt} $.

The Lie--Trotter--Suzuki formulas approximate time evolution operator $  U  $ for $H=\sum_{j=1}^m H_{j}$ as a product of the form
$$
  e^{-iHt}\approx(\prod_{j=1}^{m} e^{-i H_{j} \frac{t}{r}})^{r}.
$$

These formulas can be  defined for any integer $\chi>0$ by \cite{w1,su}
\begin{align}
S_1(t)&=\prod_{j=1}^m e^{-i H_j t/2}\prod_{j=m}^1 e^{-i H_j t/2},\nonumber\\
S_\chi(t)&=\left(S_{\chi-1}(s_{\chi-1}t)\right)^2S_{\chi-1}([1-4s_{\chi-1}]t)\left(S_{\chi-1}(s_{\chi-1}t)\right)^2\label{eq:12},
\end{align}
where $s_{\chi-1}=(4-4^{1/(2\chi-1)})^{-1}$ for any integer $\chi>1$.
This choice of $s_{\chi-1}$ is made to ensure that the Taylor series of $S_\chi$ matches that of $e^{-iHt}$ to $O(t^{2\chi +1})$.
With the values of $\chi$ large enough and the values of $t$ small enough, the approximation of $U(t)$ can reach arbitrary accuracy.

Childs et al have simulated $U(t)$ using $r$ iterations of $M_{k,k}(t/r)$ for some sufficiently large $r$ \cite{w1}:
\begin{equation}
M_{k,k}(t)=\sum_{q=1}^{k+1} C_q S_k(t/\ell_q)^{\ell_q},\label{eq:mkdef}
\end{equation}
where $\ell_q $ $(q\in\lbrace1,2,\ldots,k+1\rbrace)$ represent distinct natural numbers and $\sum_{q=1}^{k+1} C_q=1$ $(C_1,\ldots,C_{k+1} \in R)$.
In \cite{w1} , the  $\ell_q $ and $ c_{i}$ are defined as
\begin{equation}
\ell_q = \begin{cases}q, &\textrm{if $q\leq k$,}\\
e^{\gamma (k+1)},& \textrm{if $q=k+1$,}\end{cases}\label{eq:lqDef}
\end{equation}
and
\begin{equation}
C_q= \begin{cases}\frac{q^2}{q^2-e^{2\gamma (k+1)}} \prod_{j\ne q}^{k} \frac{q^2}{q^2-j^2},&\text{if $q\le k$,}\\ \prod_{j=1}^{k} \frac{e^{2\gamma (k+1)}}{e^{2\gamma (k+1)}-j^2},& \text{if $q=k+1$}.\end{cases} \label{eq:cq}
\end{equation}
The formula is accurate to $O(t^{4k+1})$ order, namely,
\begin{equation}
\norm{M_{k,k}(\lambda)-U(\lambda)} \in O(t^{4k+1}).
\end{equation}

The basic idea of this simulation algorithm is that dividing evolution time $ t $ into $ r $ segments and approximating each time evolution operator segment $ U(t/r)=e^{iHt/r} $ by a sum of multi-product formula, namely,
\begin{equation}
U(t/r)\approx M_{k,k}(t/r) =\sum_{q=1}^{k+1} C_q S_k(t/\ell_q r)^{\ell_q}.
\end{equation}

Now, we give a duality quantum computer description of the implementation of this simulation  algorithm of time evolution. The quantum circuit is  the same as Fig. \ref{f2}.
According to \eqref{eq:12}, $S_k(t/\ell_q r)^{\ell_q}  $ is an unitary operation. Let $U_{i}=S_k(t/\ell_q r)^{\ell_q}  $ and $ c_i =C_q $, $ M_{k,k}(t/r)  $ can be rewritten as
\begin{equation}
 M_{k,k}(t/r)= \sum_{i=0}^{k}c_{i}U_{i}
\end{equation}
where $U_{i}$ is an unitary operation.

It is obvious that $  M_{k,k}(t/r) $ is a duality quantum gate.
The QWD is simulated by the unitary operation $V$ and
the QWC is simulates by  unitary operation $W$
on a qudit.  The auxiliary qudit controlled operations is $U_{i}$. The matrix element $V_{i0}$ of the unitary matrix $ V $ and  the matrix element $ W_{i0} $ of the unitary matrix $ W $ satisfy:

\begin{align}\label{max}
V_{i,0}=\frac{v_{i,0}}{\sqrt{\sum_{i} |v_{i,0}|^2}}, \\
W_{0,i}=\frac{w_{0,i}}{\sqrt{\sum_{i} |w_{0,i}|^2}}.
\end{align}
As defined in \eqref{eqvw}, $ c_i $ is the product of two unitary matrix elements:

 \begin{eqnarray}
   c_i=W_{0i}V_{i0}=\frac{v_{i,0}w_{0,i}}{\sqrt{\sum_{i} |v_{i,0}|^2|w_{0,i}|^2}}.
   \end{eqnarray}
  The sum of $c_i$'s is
\begin{eqnarray}
\sum_i c_i=\sum_i W_{0i}V_{i0}=(WV)_{00}. \label{eqvw2}
\end{eqnarray}

In the special case $W=V^{\dagger}$, the simulation algorithm has the maximum success probability. The expression of  $ V_{i,0 }$ and $ W_{0,i} $  can be simplified into the form:

\begin{align}\label{maxs}
V_{i,0}=W_{0,i}=\frac{\sqrt{c_i}}{\sqrt{\sum_{i} |c_{i}|}}.
\end{align}

After implementing the QWD operation, the auxiliary qudit controlled operations and the QWC operation, detecting  the final wave function when the auxiliary qudit is in state $|0\ket$. The initial state $ |\Psi\ket|0\ket $ has been transformed into
\begin{eqnarray}
\sum_{i} W_{0i}V_{i0}U_i|\Psi\ket|0\ket=\sum_{i}
(W_{0i}V_{i0})U_i|\Psi\ket|0\ket=\sum_{i=0}^{k}c_{i}U_{i}|\Psi\ket|0\ket.
\end{eqnarray}

The approximated evolution operator  $  M_{k,k}(t/r) $ is  implemented successfully by the duality quantum computer. Implementing $ r $ segments of $  M_{k,k}(t/r) $, we can get the approximation of $ U(t) $ by $  M_{k,k}(t/r)^{r} $. Thus, this algorithm is clearly realized by the duality quantum computer in straightforward way. The essential idea of this algorithm is an iterated approximation, with each controlled $ U_{i} $ adding an additional high order approximation to the  evolution operator.

\section{Description of Berry-Childs  quantum algorithm with exponential improved precision for a sparse Hamiltonian system }\label{s6}

 Berry and Childs provided a quantum algorithm for simulating Hamiltonian dynamics  by approximating the truncated Taylor series of the evolution operator on a quantum computer \cite{Berry}.  This method is based on linear combinations of unitary operations and it can simulate the time evolution of a class of physical systems.  The performance of this  algorithm has exponential improvement over previous approaches in precision.

 Hamiltonian can be decomposed into a  linear combinations of unitary operations:
  \begin{align}\label{eq:sum}
H = \sum_{\ell=1}^L \alpha_\ell H_\ell.
\end{align}
 Dividing the evolution time into $r$ segments of length $t/r$.
 The time evolution operator of each segment can be approximated as
\begin{equation}\label{eq:truncated}
U_r := \exp(-iHt/r) \approx \sum_{k=0}^{K} \frac{(-iHt/r)^k}{k!}  ,
\end{equation}
where the Taylor series is accurate to  order $K$. Substituting the Hamiltonian in terms of a sum of $ H_\ell $ into \eqref{eq:truncated}, we can rewrite the truncated Taylor series  as \cite{Berry}
\begin{align}
U_r \approx \widetilde U &=\sum_{k=0}^{K}
\sum_{\ell_1,\ldots,\ell_k=1}^L
\frac{(-it/r)^k}{k!}\alpha_{\ell_1} \cdots \alpha_{\ell_k} \, H_{\ell_1} \cdots H_{\ell_k}. \label{eq:explict}
\end{align}

For convenience,  we can set each $\alpha_\ell > 0$. Considering $ H_\ell $ is an unitary operation, we can conclude that the approximation $ \widetilde U $  is a linear combinations of unitary operations. The expression has a quantum duality gate form.
The truncated Taylor series index can be defined as \cite{Berry}
\begin{align}
J := \{(k,\ell_1,\dots,\ell_k) \in J : k \le K\}.
\end{align}
Then, the  expression of  $ \widetilde U $  can be simplified as
\begin{align}
\widetilde U = \sum_{j \in  J} \beta_j V_j,
\end{align}
where
$\beta_{(k,\ell_1,\dots,\ell_k)} := [(t/r)^k / k!]\alpha_{\ell_1} \cdots\alpha_{\ell_k}$ and $V_{(k,\ell_1,\dots,\ell_k)} := (-i)^k H_{\ell_1} \cdots H_{\ell_k}$.
It should be noted that $ \widetilde U $ is not normalized.

We define the normalization constant as $s=\sum_{j \in  J} \beta_j $. According to
\eqref{e1}, $ L_r =\widetilde U/s$ is a quantum duality gate. We let $ r_i=\beta_i/s $ , $ V_j= U_i $, then it comes back to the duality quantum gate form in \eqref{e3},

\begin{eqnarray}
L_r=\sum_{i=0}^{d-1} r_i U_i,
\end{eqnarray}
where $ i\in\lbrace 0, 1, 2, \ldots, d-1\rbrace $.

To give the duality quantum computer description, we need  to realize the following processing
\begin{eqnarray}
|\Psi\ket|0\ket\rightarrow |\Psi\ket
\widetilde U|0\ket.
\end{eqnarray}
\begin{figure}
\begin{center}
\includegraphics[width=9cm,angle=0]{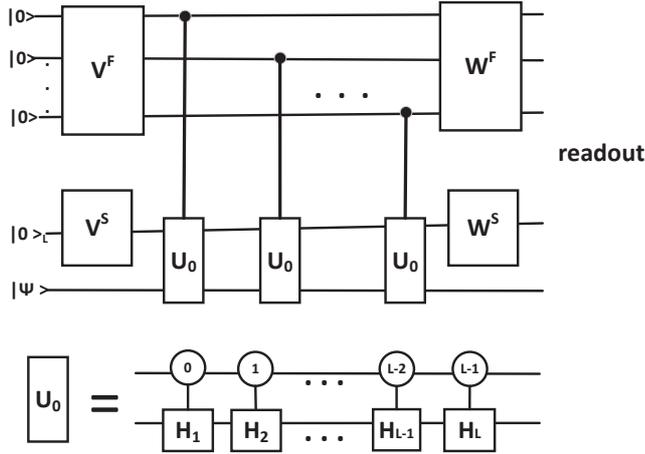}
\caption{Duality computing in a quantum computer.  $|\Psi\rangle$ is
the initial state of duality quantum computer and there are $ K $numbers of $|0\rangle$ auxiliary controlling qubit and an $ L$ level auxiliary qudit$ |0\ket_{L} $ auxiliary controlling qudit.  Unitary operations $U_{0}$ are activated only when
the qubit and qudit holds the respective values indicated in circles. The each unitary operation $U_{0}$ is composed of $ H_{1}, H_{2}, \ldots, H_{L-1}, H_{L}$.  \label{f3}}
\end{center}
\end{figure}
In Fig. \ref{f3}, we give an illustration for our method to perform the algorithm in the form of  quantum circuit. The unitary operation $U_{0}$ corresponds to the decomposing form of  Hamiltonians: $ H = \sum_{\ell=1}^L \alpha_\ell H_\ell $ and the quantum circuit In Fig. \ref{f3}  implements $ U_r = \exp(-iHt/r) \approx \sum_{k=0}^{K} \frac{(-iHt/r)^k}{k!}$.

The implementation of operation $ \widetilde U $ need  an $ L$ level auxiliary qudit $ |0\ket_{L} $ and $ K $  auxiliary qubits $ |0\ket^{K} $ which correspond to  implementation of two QWD   operations and QWC operations. Actually, the equation of \eqref{eq:explict} indicates that we need  summarize twice to realize the right side of this equation. We express the initial state as
$|\Psi\ket|0\ket^{K}|0\ket_{L}$

Firstly, we transform the $|0\ket^{K}$ part of the initial state into the normalized state using the QWD operation. We have
\begin{equation}
|0\ket^{K} \rightarrow \sum_{k=0}^K \sqrt{t^k/k!}|{1^k0^{K-k}}\ket.
\end{equation}
 We let $f= \sum_{0}^{K}\frac{t^{k}}{k!}$ and define the QWD as $ V^{F} $, which  can be expressed as a $2^{K} \times 2^{K}$ matrix.
The elements of the matrix satisfy
\begin{align}
V^{F}_{i,0}=\frac{v^{F}_{i,0}}{\sqrt{\sum_{i} |v^{F}_{i,0}|^2}}, \label{eq:firstd}
\end{align}
where
\begin{eqnarray}
v^{F}_{i,0}=\left\{
\begin{array}{rcl}
\sqrt{\frac{t^{k}}{k!}}, &  & {i=2^{K} -2^{K-k}},k\in \lbrace 0,1,\ldots ,K \rbrace. \\
 0   ,                             & & {\text{else.}}
\end{array} \right.
\end{eqnarray}
After implementing the first unitary operations  $ V^{F} $ in the $ |0\ket^{K} $ part of the initial state, we can get normalized state $\frac{\sum_{k=0}^K \sqrt{t^k/k!}}{\sqrt{f}}|{1^k0^{K-k}}\ket$.

Secondly, using the QWD operation once again to transform the $ |0\ket_{L} $ part of initial state into the normalized state $\sum_{\ell=1}^L \sqrt{\alpha_{\ell}}|{\ell}\ket$. We let $g= \sum_{\ell=1}^L \alpha_{\ell} $ and define the second QWD operation as $ V^{S} $, which  can be expressed as a $L \times L $ matrix.
The elements of the matrix satisfy
\begin{align}
V^{S}_{\ell,0}=\frac{v^{S}_{\ell,0}}{\sqrt{\sum_{\ell} |v^{S}_{\ell,0}|^2}}, \label{eq:second1}
\end{align}
where
\begin{align}
v^{S}_{\ell,0}=\sqrt{\alpha_{\ell}}.  \label{eq:second2}
\end{align}
After implementing the second unitary operations  $ V^{S} $ on $ |0\ket_{L} $ part of initial state, we can get normalized state $\sum_{\ell=1}^L \frac{\sqrt{\alpha_{\ell}}}{\sqrt{g}}|\ell\ket$.

We perform the  $ L$ level auxiliary qudit $ |0\ket_{L} $ and $ K $  auxiliary qubits $ |0\ket^{K} $ controlled operation $ U_{i} $ on the computer. The processing can be described as
\begin{align}
|\Psi\ket\rightarrow \sum_{k=0}^K \sqrt{t^k/k!} \sum_{\ell=1}^L \alpha_{\ell}U_{i}|\Psi\ket.  \label{eq:second3}
\end{align}

Then, corresponding to the two times of  performing of the  QWC  operations, we need  perform the QWC  operations  twice to combine the wave functions.  We define the QWC operations   $ W^{F} $and $W^{S} $  corresponding to the QWD  operations $ V^{F} $ and $ V^{S} $ respectively.

We set  $ W^{F} =(V^{F} )^{\dagger}$ , $W^{S}=( V^{S})^{\dagger} $ and
perform the the QWC operations $ W^{F} $and $W^{S} $ on the state $|{1^k0^{K-k}}\ket$ and $|\ell\ket$, respectively.

After  QWC operations $ W^{F} $and $W^{S} $, we detect the final wave function when the auxiliary system is in state
$|0\ket^{K}|0\ket_{L}  $. In the final state, we only focus our attention on the terms  with the $ L$ level auxiliary qudit in state $ |0\ket_{L} $ and $ K $  auxiliary qubits in state $ |0\ket^{K} $. We have the following
 \begin{eqnarray}
&&{1\over \sqrt{f}}\sum_{k=0}^K \sqrt{t^k/k!}|{1^k0^{K-k}}\ket\rightarrow {1\over f}\sum_{k=0}^Kt^k/k!}|{0\ket^{K}, \\
&&\sum_{\ell=1}^L \frac{\sqrt{\alpha_{\ell}}}{\sqrt{g}}|\ell\ket\rightarrow {1\over g}\sum_{\ell=1}^L \alpha_{\ell}|0\ket_{L}.
\label{eq:com1}
 \end{eqnarray}

 It should be noted that the summation parts $\sum_{k=0}^K \sqrt{t^k/k!} $ and $\sum_{\ell=1}^L \alpha_{\ell} $ already have been combined with $ U_{i} $.
The initial state is transformed into
\begin{align}
|\Psi\ket|0\ket^{K}|0\ket_{L} \rightarrow {1\over fg}\sum_{k=0}^Kt^k/k!\sum_{\ell=1}^L \alpha_{\ell}U_{i}|\Psi\ket|0\ket^{K}|0\ket_{L},  \label{eq:com3}
\end{align}
where $ U_{i} $ corresponds to some $(-i)^k H_{\ell_1} \cdots H_{\ell_k}$ and $ i\in J $.

It is obvious that $ s= fg$ and $  \sum_{j \in  J} \beta_j V_j=\sum_{k=0}^Kt^k/k!\sum_{\ell=1}^L\alpha_{\ell} $. Consequently,  we have successfully realized the following process:

\begin{align}
|\Psi\ket|0\ket^{K}|0\ket_{L} \rightarrow  \dfrac{1 }{s}\widetilde U|\Psi\ket|0\ket^{K}|0\ket_{L} .  \label{eq:com3}
\end{align}

Finally, the robust form of obvious amplitude amplification procedure of  \cite{Berry} enables us to deterministically implement $ \widetilde U $ through amplifying the amplitude of $ \dfrac{1 }{s}\widetilde U $.  The approximation accuracy of  $ \widetilde U   $ can be quantified by approximation error $ \epsilon $.  According to the Chernoff bound, as studied in \cite{Berry2}, the query complexity is
\begin{equation}
K = O\biggl( \frac{\log(r/\epsilon)}{\log\log(r/\epsilon)} \biggr),
\end{equation}
and the error of approximation in each segments satisfy:
\begin{equation}
\parallel \widetilde U-U_{r}\parallel \leq \frac{\epsilon}{r}.
\end{equation}

The total number of gates in the simulation for time $t/r$ in each segment is \cite{Berry}
\begin{align}
\label{gate}
O\biggl(\frac{L(n + \log L)\log(T/\epsilon)}{\log\log(T/\epsilon)}\biggr).
\end{align}
In the duality quantum computer description, our method gives a slight improvement than \cite{Berry}, which uses
\begin{align}
\label{wgate}
O\biggl(\frac{L(n + 1)\log(T/\epsilon)}{\log\log(T/\epsilon)}\biggr)
\end{align}
gates.

Thus, we have given  a standard program in the duality quantum computer to realize the simulation methods of  Hamiltonians based on linear combinations of unitary operations. The physical picture of our description of the algorithm is clear and simple: each of QWD and QWC operations will lead to one summation of linear combinations of unitary operations. Our method is intuitive and can be easily performed based on the form of time evolution operator.

\section{Summary} \label{s7}

In the present paper, we have briefly reviewed the duality quantum computer. Quantum wave can be divided and recombined by the QWD  and QWC operations in a duality quantum computer. The divider and combiner operations are two crucial elements of
operations in duality quantum computing and they are realized in a quantum
computer by unitary operators.
Between the dividing and combining operations, different computing gate operations can be performed at the different sub-wave paths which is called the duality parallelism.
It enables us to perform linear combinations of unitary
operations in the computation, which is called the duality quantum gates or the generalized quantum gates. The duality parallelism  may exceed quantum parallelism in quantum computer in the precision aspect.

The duality quantum computer can be perfectly simulated by an ordinary quantum computer with $n$-qubit and an additional qudit, where a qudit labels the slits of the duality quantum computer. It has been shown that the duality quantum computer is able to simulate any linear bounded operator in a Hilbert space, and unitary operators are just the extreme points of the set of generalized quantum gates.

Simulating the time evolution of quantum systems or the dynamics of quantum systems is a major potential application of quantum computers. The property of duality  parallelism  enables duality quantum computer to simulate the dynamics of  quantum systems using linear combinations of unitary operations.
It is naturally suitable to realize the simulation algorithms of Hamiltonians based on multi-product formulas which are usually adopted in  classical algorithms. The duality quantum computer can be used as a bridge to transform classical algorithms into quantum computing algorithms. We have realized both Childs-Wiebe algorithm  and Berry-Childs simulation algorithms in a duality quantum computer. We showed that their algorithm can be described straightforwardly in a duality quantum computer.  Our method is simple and has a clearly physical picture. Consequently, it can be more easily realized in experiment \cite{flong,flong2}.

This work was supported by the National Natural Science Foundation
of China Grant Nos.  (11175094 , 91221205), the National Basic
Research Program of China (2011CB9216002), the Specialized Research
Fund for the Doctoral Program of Education Ministry of China.

%33333333333333333333333333333333333333333333333333333333333333333333333333

\end{document}